# Decoherent Focusing Design and Advanced Passive Speckle Reduction for Laser Illumination Systems

**Anatoliy Lapchuk[1*], Olexandr Prygun[1], Dmytro Manko[1], Ivan Gorbov[1,2], and Yevhenii Morozov[3*]**

[1] Department of Optical Engineering, Institute for Information Recording of NAS of Ukraine, Kyiv, Ukraine
[2] Optics and Photonics Research Group, Faculty of Engineering, University of Nottingham, Nottingham, UK
[3] AIT-Austrian Institute of Technology, Vienna, Austria

*E-mails: alapchuk@yahoo.com; yevhenii.morozov@ait.ac.at

## Abstract

To effectively apply passive speckle reduction methods, it is essential to use an illumination system that maximally exploits the non-ideal temporal coherence and angular diversity (spatial coherence reduction) of laser light. This study examines the necessary conditions for these factors to act independently to achieve maximum speckle reduction. A novel design for decoherent focusing of laser illumination into a rectangular, uniformly lit spot is proposed. The design is based on an array of rectangular prisms of varying heights and a Fresnel lens comprising crossed 1D cylindrical Fresnel lenses with flat facets. Mathematical modelling demonstrates that such a lens can effectively focus a Gaussian beam into a rectangular uniform illuminated spot, even for lenses with a high numerical aperture ($NA = 0.2$). A planar implementation of this lens is also proposed. The comb-like spectrum of a laser diode is shown to limit the capabilities of passive methods, as it is challenging to generate a significant number of decorrelated laser sub-beams beyond the available spectral modes. Experimental results confirm that applying a slight modulation (~15%) to the laser diode current during intensity integration transforms the comb spectrum into a continuous one. This enables the generation of an unlimited number of decorrelated laser beams, independent of the laser's spectral modes, to achieve the desired number for a specific optical device.

Keywords: speckle suppression, laser projector, decoherence, decoherent focusing system, passive system

## 1. Introduction

Lasers are highly efficient and the brightest sources of light, emitting highly collimated and coherent beams [1, 2]. This unique property enables the creation of images with the most vivid colours and the widest colour gamut [3]. High-power multimode laser diodes, commonly used in laser projectors, exhibit a complex intensity distribution across their cross-section [4], which also varies with the applied current [5]. This unstable and intricate intensity distribution must be efficiently transformed by the illumination system of a laser projector into a rectangular, uniformly lit spot. Due to the unique light distribution across the beam cross-section of each laser, it is most practical to employ a universal method for achieving uniform illumination. Such a method should function effectively regardless of the laser type or individual sample characteristics. Universal approaches direct light from various sections of the laser beam into a rectangular screen area, averaging the cross-sectional distribution and creating uniform illumination. This can only be achieved when light from different sections of the beam reaches the screen at varying angles. However, due to the coherence of the laser beam, such uniform illumination is often significantly distorted by interference effects. The most widely used system employing this approach consists of two arrays of rectangular microlenses and a Fourier lens (**Figure 1a**). The Fourier lens focuses the light into a rectangular area, producing uniform illumination [6, 7]. **Figure 1bc** illustrates the results of such a universal illumination system for both incoherent and coherent laser light [8]. In the case of laser illumination, instead of uniform lighting, a rectangular grid of light spots with equal intensity is observed (**Figure 1c**), replacing the uniformly illuminated spot seen in the case of incoherent illumination (**Figure 1b**). To achieve uniform illumination with coherent light, an optical system is required that can render the beams arriving at different angles incoherent. This would effectively eliminate interference effects and produce a uniformly lit area [8].

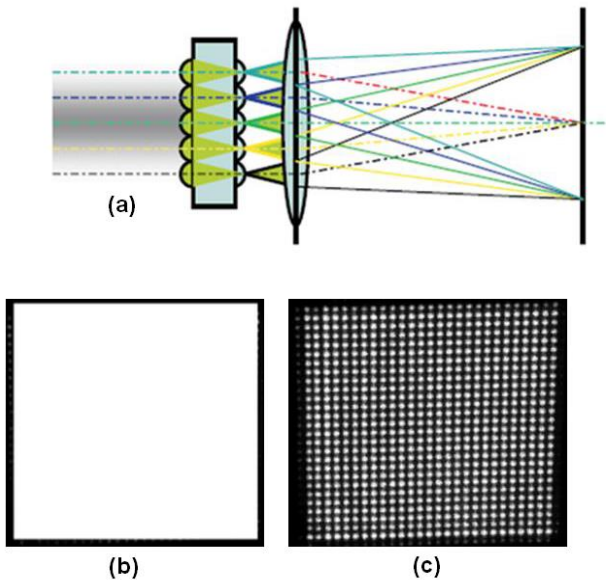

**Figure 1.** (a) Optical system for uniform illumination without significant optical energy loss; (b) optical spot under incoherent illumination; (c) optical spot under laser illumination. Reprinted from [8].

Another significant drawback of using coherent light sources in imaging systems is the appearance of granular modulation in perceived images, known as subjective speckles, which arise from light scattering on the screen [9]. The intensity of speckle noise is determined by the speckle contrast $C$:

$$C = \sigma / \bar{I} , \qquad (1)$$

where $\bar{I}$ is the average intensity of the uniformly illuminated screen in the generated image, and $\sigma$ is the standard deviation of the intensity. **Figure 2** illustrates the impact of speckles on intensity modulation and the distortion of the colour gamut in the generated image caused by speckles [10]. Subjective speckles are also a result of the interference of light scattered on the screen within the eye [9]. Therefore, to use laser illumination, the optical system must incorporate a mechanism for speckle reduction, utilizing one method or another.

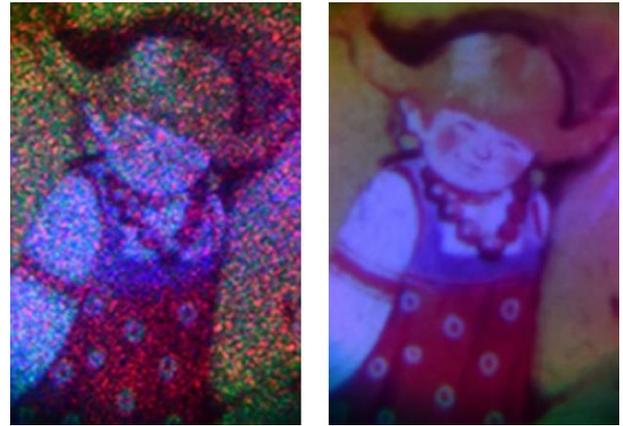

**Figure 2.** The image generated with coherent laser illumination (left) and decoherent illumination (right). Reprinted from [10] with permission from Elsevier.

For characterizing a speckle reduction method, the speckle reduction coefficient $k = C_0 / C$ is used, where $C_0$ and $C$ are the speckle contrasts before and after applying the speckle reduction method, respectively. Speckle reduction in the eye can be achieved by averaging decorrelated speckle patterns, which can be generated by decorrelating portions of light scattered on the screen. When an image is created using multiple decorrelated speckle patterns (assuming that all independent speckle patterns have the same speckle contrast), the speckle reduction coefficient for the method is calculated using the formula [9]:

$$k = \sum_{i=1}^{N} I_i / sqrt\left( \sum_{i=1}^{N} I_i^2 \right) , \qquad (2)$$

where $I_i$ is the intensity of the $i$-th speckle pattern and $N$ is the number of decorrelated speckle patterns. Under the condition of equal intensity of each speckle patterns, Eq. (2) gives $k = \sqrt{N}$ . The creation of decorrelated speckle patterns can be achieved by reducing the temporal coherence of laser illumination [4], where the laser light exhibits a short coherence length. This effect is achieved through the screen's surface relief, where the scattered light has optical path differences exceeding the coherence length of the laser beam. As a result, the light forms decorrelated speckle patterns. The speckle reduction coefficient for this method is calculated using the formula provided in [9]:

$$k = \left(1 + 8\pi^2 (\Delta \lambda / \lambda)^2 (\Delta h / \lambda)^2\right)^{1/4}, \qquad (3)$$

where $\Delta h$ is the standard deviation of the optical path length of the light scattered on the screen, $\Delta \lambda$ is the spectral width of the laser, and $\lambda$ is the central wavelength of the laser radiation. A significant effect is achieved when the coherence length is much smaller than the standard deviation of the optical path length of the light scattered on the screen. This occurs when the spectral width is on the order of several nanometres. In this case, the speckle reduction coefficient can be estimated using the formula:

$$k \approx \left(2\pi \sqrt{2} (\Delta \lambda \cdot \Delta h / \lambda^2)\right)^{1/2} = (N)^{1/2}, \qquad (4)$$
$$= (s \cdot \Delta \lambda / \lambda^2)^{1/2}$$

where $s = 2\pi \sqrt{2} \cdot \Delta h$. Since the spectral width of laser systems does not normally exceed 2 nm, this method achieves only a modest speckle reduction, typically by a factor of a few. This level of reduction is insufficient for laser projectors. The speckle reduction effect can also be achieved through depolarization of the illumination, which involves combining two orthogonal polarizations of laser light [11]. However, since light can have only two independent polarizations, the speckle reduction coefficient for this method does not exceed $\sqrt{2}$.

The method based on spatial decorrelation or angular diversity decorrelates light beams incident on the screen at different angles to generate decorrelated speckle patterns and reduce speckle noise. The creation of decorrelated beams arriving at the screen from varying angles is also necessary for achieving uniform illumination; however, there is a significant distinction. To achieve decorrelation of speckle patterns, the difference in incidence angles between the beams on the screen must be greater than the angular resolution of the eye relative to the screen [9]. Therefore, to achieve a substantial speckle reduction effect, the optical system must produce a beam with a significant étendue $G$ [12]:

$$G = S_0 \cdot \Omega, \qquad (5)$$

where $S_0$ is the cross-sectional area of the conical beam at the waist and $\Omega$ is the solid angle of beam divergence.

Multimode lasers with transverse modes, due to their small size of just a few microns, have low $G$, and conventional optical systems based on lenses do not alter the $G$. Therefore, a straightforward reduction of the spatial coherence of lasers has little impact on the ability to reduce speckle levels in laser projectors, particularly due to the large screen sizes involved. However, owing to the significantly larger diameter of the projector's objective lens compared to the diameter of the human eye, angular diversity can generate up to $(NA_{obj}/NA_{eye})^2$ where $NA_{obj}$ and $NA_{eye}$ are the numerical aperture of the objective lens and human eye respectively, or several hundred, decorrelated speckle patterns. This is the primary mechanism for achieving substantial speckle reduction.

From the above it follows that the challenge of creating uniform lighting and speckle-free lighting shares a common origin – namely, the phenomenon of interference. Consequently, they may leverage a shared mechanism to achieve high-quality lighting. Therefore, it is evident that an optimal optical system should be designed to fully utilize this principle.

## 2. Temporal and angular decorrelation of light beams during active and passive angular decorrelation of sub-beams with varying angles of incidence

Ideally coherent illumination produces a speckle contrast of 0.71 upon scattering on a screen, whereas achieving high-quality imaging requires the speckle contrast to not exceed 0.04 [13, 14]. This necessitates a significant reduction in speckle contrast—approximately 18-fold. Given the need for such a substantial reduction, it is crucial to employ all available methods for minimizing speckle contrast, including reducing temporal, spatial, and polarization coherence. Furthermore, these methods must generate decorrelated speckle structures independently of one another. In this case, the overall speckle reduction factor will be the product of the reduction factors achieved by decreasing the temporal, spatial, and polarization coherence of the laser illumination:

$$k_{tot} = k_t \cdot k_\Omega \cdot k_p, \qquad (6)$$

where $k_{tot}$ is the total speckle reduction coefficient, and $k_t$, $k_\Omega$, and $k_p$ are speckle reduction coefficient resulting from the reduction of temporal, spatial, and polarization coherence of the illumination, respectively.

Decorrelation of laser beams can be achieved using either active or passive methods. Active methods employ optical elements that dynamically alter the phase or propagation direction of laser beams to decorrelate speckle structures. This can be accomplished using a liquid crystal matrix [15, 16] or by mechanically shifting a diffractive optical element (DOE). However, the minimum switching time of a liquid crystal panel is approximately 0.4 ms [17], which limits the maximum phase-switching frequency of liquid crystal layers to a few hundred Hz. This frequency is insufficient to reduce speckles to the required level (~10 kHz). In contrast, mechanical shifting of a diffractive optical element (DOE) results in varying phase-shifting speeds for different DOE diffraction orders. This enables the decorrelation of diffraction orders and provides angular diversity, which aids in speckle reduction [18, 19]. To effectively reduce speckles, a DOE is required that produces numerous diffraction orders, with these orders uniformly filling the angular cone of the projector objective's numerical aperture. In previous works, a method based on 2D DOE structures was proposed and developed, utilizing pseudorandom sequences. These sequences enable the generation of a sufficient number of diffraction orders, whose decorrelation can be achieved through simple linear shifting of the DOE [10, 20–22]. However, achieving significant speckle reduction with DOEs requires a large period (or pixel size), which increases the size of the projection system [23] and demands high shifting speeds [19]. Furthermore, this method does not leverage angular decorrelation of laser sub-beams to achieve uniform illumination. These limitations prevent the development of a compact and technologically advantageous design for a projector's illumination system based on this approach.

In passive methods utilizing angular diversity, the beam is divided into sub-beams with path differences greater than the coherence length of the laser beam. These sub-beams are then directed onto a screen at angles with sufficient angular differences to produce decorrelated speckle patterns. Known implementations of this method include long optical multimode fibres [24, 25] and rectangular multimode lightguides [26, 27]. However, both approaches require active optical elements to achieve the speckle reduction necessary for laser projectors. Additionally, they are not compact and suffer significant optical losses due to multiple reflections, making them suboptimal engineering solutions for the problem.

In [8], the potential for developing a passive, compact laser system for uniform illumination is discussed, but without a detailed analysis of the parameters or sufficient theoretical evaluation of the method. Below, we will provide a more in-depth examination of the relationship between temporal and spatial depolarization in such a system, propose a new implementation of the method, and analyse the feasibility of achieving compact, uniform illumination through mathematical modelling. Additionally, an evaluation of the speckle reduction efficiency using this method will be conducted based on the analysis of experimental data.

**Figure 3** illustrates an optical scheme for scattering sub-beams from a screen, employing methods that combine temporal partial incoherence and angular decorrelation of sub-beams to suppress speckle.

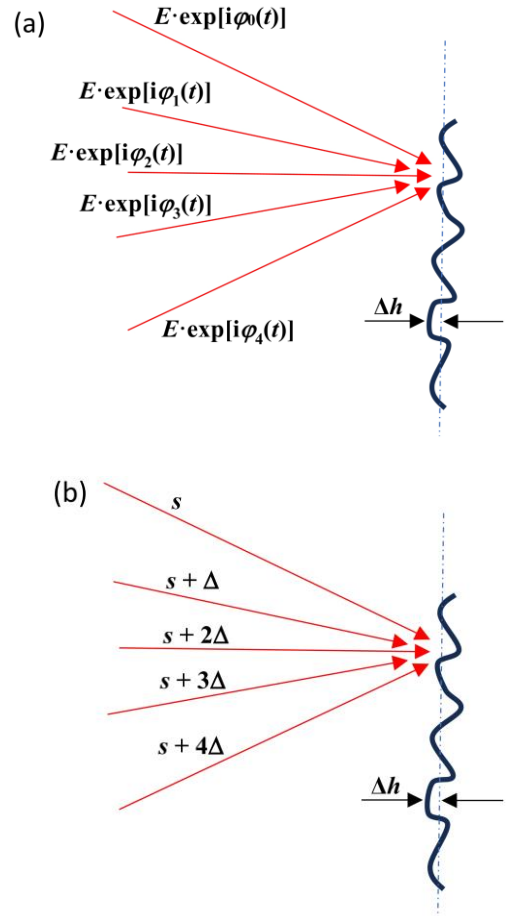

**Figure 3.** Schematic of the light scattering on a screen by sub-beams at angles decorrelated for speckle patterns for (a) the active sub-beam decorrelation method: $\langle \exp[i\phi_n(t)]\exp[i\varphi_m(t)] \rangle = \delta_{nm}$ – averaging over the integration time of intensity in the eye, where $\delta_{nm}$ is the Kronecker delta; and (b) the passive method.

In these optical setups, it is assumed that all sub-beams have angular differences larger than the angular resolution of the human eye, ensuring that they create decorrelated speckle patterns if they are indeed decorrelated. In the active method, sub-beams are decorrelated through two mechanisms. The first mechanism involves active phase modulation, achieved using active elements such as shifting a diffractive optical element (DOE), moving a diffuser, or introducing a time delay between sub-beams incident on the screen at different angles using a vibrating mirror (**Figure 3a**). The second mechanism leverages partial temporal incoherence, which suppresses speckle by introducing differences in the optical path lengths of light scattered from the screen, exceeding the coherence length of the laser beam. These two methods rely on distinct mechanisms for sub-beam decorrelation. Sub-beams generated through both mechanisms are decorrelated, and the overall speckle suppression coefficient is the product of the suppression coefficients for temporal incoherence and spatial coherence reduction.

In the passive method (**Figure 3b**), the optical system divides a laser beam into several sub-beams, which are directed onto the screen via optical paths of varying lengths sufficient to achieve sub-beam decorrelation. To achieve this, a long multimode optical fibre [24, 25] or a light tube [26, 27] can be employed in multimode operation to generate decorrelated sub-beams. Alternatively, a multiretarder plate, consisting of an array of rectangular prisms with varying heights, can be used to create sub-beams separated either in angular or real space. In the passive method, angular diversity and partial temporal incoherence rely on differences in optical path lengths to decorrelate the sub-beams, utilizing the same underlying mechanism for sub-beam decorrelation. However, it can be shown that simply decorrelating sub-beams incident on the screen at respective angles is not a sufficient condition for optimal speckle reduction. In this case, the speckle suppression coefficient may be lower than the product of the suppression coefficients resulting from reductions in spatial and temporal coherence of the illumination [28]. To clarify this, let us assume that a set of the sub-beams produced by the optical system have optical path differences that are multiples of $\Delta = n_0 \cdot l$, where $n_0$ is an integer larger or equal to 1, and the optical path difference of light scattered on the screen is equal to $\nu \cdot l$ (see **Figure 3**), where $\nu \approx 2\pi\sqrt{2} \cdot \Delta h / \left(\lambda^2 / \Delta\lambda\right)$ and $l = \lambda^2/\Delta\lambda$ is the decorrelation (coherence) length of the laser beam.

The transformation of laser illumination along the optical path during its propagation through the illumination system with a passive speckle reduction method (see **Figure 4**) can be described by the following formula:

$$E_0(i) \Rightarrow \sum_{i=0}^{N-1} E_i(s+in_0 \cdot l) \Rightarrow \sum_{i=0}^{N-1}\sum_{j=0}^{\nu-1} e_{ij}(s+in_0 \cdot l + j \cdot l) \; . \; (7)$$

At the output of the optical scheme, we have $N \cdot \nu$ sub-beams, however we are specifically interested in the number of decorrelated sub-beams $N_0$:

$$N_0 = \begin{cases} (N-1) \cdot n_0 + \nu, \; n_0 < \nu \\ N \cdot \nu, \; n_0 \geq \nu \end{cases} . \quad (8)$$

Thus, applying this method will yield a speckle reduction coefficient (a majoritarian estimate based on the number of decorrelated sub-beams, without accounting for their varying intensities):

$$k = \begin{cases} sqrt\left(\left(k_\Omega^2 - 1\right)n_0 + k_t^2\right), \; n_0 < \nu \\ k_\Omega \cdot k_t, \; n_0 \geq \nu \end{cases} , \quad (9)$$

where $k_\Omega = sqrt(N)$, $k_t = sqrt(\nu)$ are the speckle reduction coefficients when applying only angular decorrelation or reducing the temporal coherence of illumination, respectively. Applying Eq. (9), it is straightforward to calculate that with $n_0 = 1$ and $\nu = 9$, the method provides no significant advantage in speckle reduction from the temporal coherence of the laser beam, yielding only about 9.5%. However, with a substantial increase in the optical path difference within the optical system, such that $n_0 \geq \nu$, both spatial and temporal coherence reductions can be fully utilized. In this case, reduced temporal coherence of the laser beam contributes an additional 67% reduction in speckle intensity. Thus, for the passive method to fully utilize both angular and reduced temporal coherence, the optical path difference of the sub-beams within the illumination system must be equal to or greater than the optical path difference generated by light scattering on the screen.

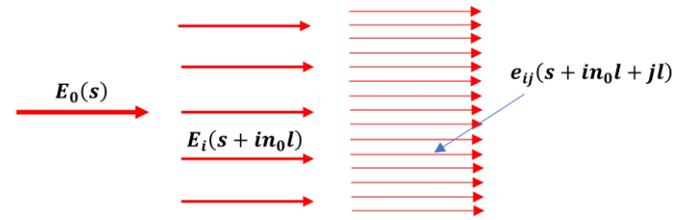

**Figure 4.** The scheme of laser illumination transformation along the optical path during its propagation through the illumination system with a passive speckle reduction method.

## 3. Optical scheme of decoherent focusing with two Fresnel cylindrical lenses for achieving uniform rectangular illumination

We proposed an optical scheme for incoherent focusing that utilizes a 2D prism array to achieve a compact rectangular illumination system [8]. In the previous study, evaluations of the optical system were conducted without performing numerical modelling to assess the uniformity of the resulting illumination. However, fabricating prisms of varying heights and angles, followed by the precise alignment of each prism along both the optical axis and the transverse cross-section, is a challenging technological process. Additionally, such a system requires the rear 1D prism-based focusing system to focus through the front 1D prism-based focusing system, which is composed of prisms with varying heights. This configuration degrades the quality of the resulting light spot. Therefore, it is essential to find a way to improve the design of incoherent focusing and make its fabrication more technologically feasible.

**Figure 5** illustrates a modification of the proposed method, which consists of a 2D array of rectangular prisms creating optical path differences $S_{ij}$, so that $\text{abs}\left(S_{ij} - S_{mn}\right)/\left(n_r - 1\right) \ge vl$, where $n_r$ is a refractive index of the prism medium, for all different prisms that is $m \ne i$ and $n \ne j$, and two Fresnel cylindrical lenses rotated by 90°. These lenses focus the laser beam into a rectangular spot by mixing light from different areas of the laser beam, similar to the approach used in the well-known design featuring two microlens arrays and a Fourier lens.

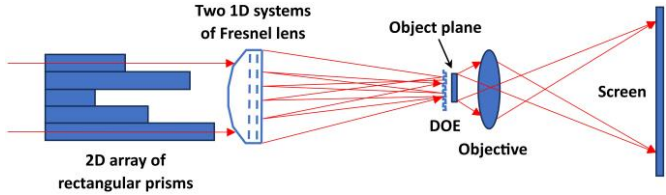

**Figure 5.** The principal optical scheme of the proposed illumination system.

A possible implementation of such a Fresnel lens design is shown in **Figure 6**. This system consists of two Fresnel lenses rotated by 90°, where the cylindrical surfaces are replaced with flat regions connecting the extreme points of the cylinder and dividing the beam into equal-width strips. The lens parameters are selected to ensure a common focal plane. This combined lens has no sharp protruding parts, making it suitable for manufacturing through hot pressing. However, it has one drawback: light passing through the first surface travels through varying thicknesses of the second lens, introducing additional aberrations to a lens that already has significant inherent aberrations. **Figure 7** shows a planar implementation of such a lens that operates using the same principle. The slopes of each region of this prismatic lens can be easily calculated using well-known formulas. This lens design exhibits fewer aberrations but is more challenging to manufacture due to the presence of sharp angles and protrusions.

For effective speckle reduction, the 2D array of rectangular prisms must have an optical path difference between the prisms greater than the optical path difference of the scattered light on the screen. The 2D array can be implemented either as a direct 2D array (**Figure 8a**) or as a sequence of two 1D arrays (**Figure 8b**) with orthogonal orientations, placed one above the other. Unlike the previous design, the order of rectangular prisms in the array does not matter, and the system is insensitive to the displacement of individual prisms along the optical axis.

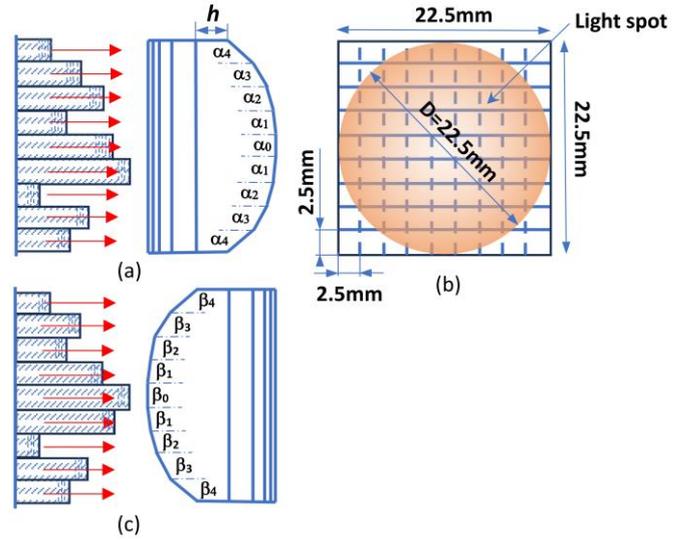

**Figure 6.** Optical system comprising two 1D cylindrical Fresnel lenses with a continuous surface profile and a 2D array of rectangular prisms of varying heights, designed for decoherent focusing of illumination into a rectangular light spot. (a) Top view; (b) Front view showing the light spot (orange circular spot); (c) Side view. Red arrows indicate the light rays.

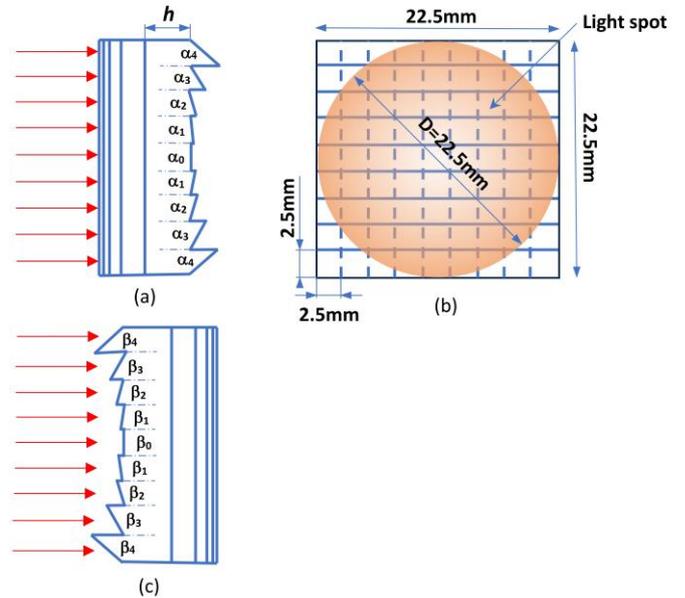

**Figure 7.** Optical system featuring two 1D Fresnel lenses in a planar configuration with a sawtooth surface profile, designed to focus illumination into a rectangular light spot using a 2D prism array. (a) Top view; (b) Front view showing the light spot (orange circle); (c) Side view. Red arrows indicate light rays.

The transverse position of the 2D array of rectangular prisms must be aligned with the transverse position of the lens so that their flat boundaries coincide. Additionally, both optical elements must have matching transverse dimensions. Additionally, it is proposed to place a 2D diffractive optical element (DOE) based on pseudorandom sequences in the focal plane of the Fresnel lens. The position of the DOE in the optical system is shifted by a few millimetres relative to the plane optically conjugated with the screen, providing additional smoothing of illumination at high spatial frequencies (see **Figure 5**).

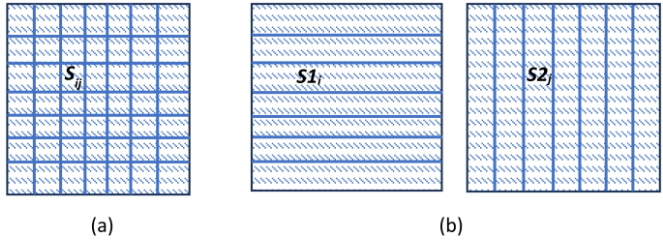

(a)                                                          (b)

**Figure 8.** (a) 2D array of rectangular prisms having different thickness, $\min\left(\text{abs}\left[S_{ij} - S_{mn}\right]/\left[n_r - 1\right] \geq \nu l\right)$, of dimension $N_1 \times N_2 = 7 \times 7$; (b) two 1D arrays of long rectangular prisms with the orthogonal orientation of arrays; first 1D array has $N_1 = 7$ elements having different thicknesses $S1_n$ and $\min\left(\text{abs}\left[S1_i - S1_m\right]/\left[n_r - 1\right] \geq \nu l\right)$ for $i \neq m$; second array has $N_2 = 7$ element having different thicknesses $S2_n$ and $\min\left(\text{abs}\left[S2_j - S2_n\right] \geq N_1 \cdot \nu l/\left[n_r - 1\right]\right)$ for $j \neq n$; $n_t$ is the refractive index of the prism medium.

## 4. Modelling of the uniform illumination system

The system of two cylindrical prisms exhibits significant aberrations. Additionally, the focusing performance of the first lens is influenced by variations in the thickness of the second lens, as the focused rays from the first lens pass through the latter. Therefore, it is necessary to verify the practical implementation of the method. At this stage, we do not have the capability to manufacture such a Fresnel lens, and due to its non-standard design, it cannot be purchased. However, this incoherent illumination system can be accurately modelled within the framework of ray optics. The SolidWorks 2020 CAD package was used to create a model of the illumination system and TracePro 7.3 for performing the ray tracing by the Monte Carlo method (see **Figure 9**). We conducted simulations of uniform illumination for a lens based on cylindrical surfaces, where a Gaussian laser beam truncated by an aperture at the 1/e level was used as input. Each 1D lens consisted of nine flat regions of equal width (2.5 mm) made from BK7 glass with a refractive index of $n = 1.518$, wavelength of the light source was set to 546.1 nm.

The light beams are colour-coded based on their intensity relative to the Gaussian beam core: red beams represent intensities ranging from 67% to 100% of the core's intensity, green beams correspond to intensities between 34% and 66%, and blue beams indicate intensities within the range of 0% to 33%. Simulations were performed for lenses with varying focal lengths. The results indicate that higher optical power results in greater aberrations, which distort the quality of the illumination. However, a lens with higher optical power allows speckle reduction over a larger screen area by creating a beam with greater étendue, while also enabling a more compact illumination system. Since the system size fits well within the assumptions of ray optics, the dimensions of the optical system – specifically, the widths of the flat regions and the focal length – do not significantly affect the simulation results.

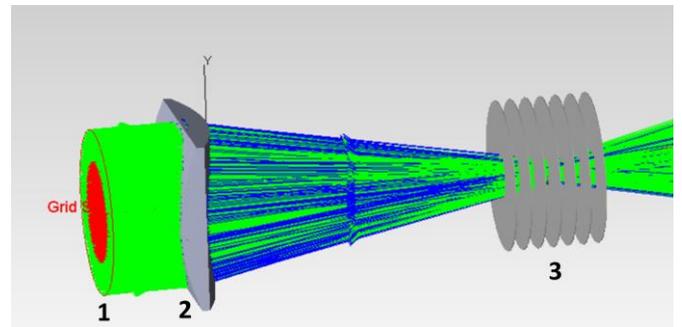

**Figure 9.** Simulation of light intensity distribution using the Monte Carlo ray-tracing method: 1 – Grid light source, 2 – the lens, and 3 – a set of fully transparent monitors.

**Figures 10 and 11** show the simulation results for two lenses with different optical powers. The parameters of the 1D cylindrical lenses for each combined optical system are listed in **Table 1**, corresponding to the long-focal lens (Lens 1, **Figure 10**) and the short-focal lens (Lens 2, **Figure 11**). The long-focal system (Lens 1) and the short-focal system (Lens 2) focus light at distances of 124 mm and 45 mm, respectively, measured from the vertex of the rear cylindrical lens. Lens 1 and Lens 2 have different thicknesses h of 4 mm and 1 mm, respectively. As seen from the results, this optical system effectively focuses the Gaussian beam into a rectangular spot with minimal optical losses. Even for Lens 2, with a higher optical power ($NA \approx 9/45 = 0.2$), the focus yields a nearly ideal rectangular spot with uniform illumination. Comparing the focusing performance of the two lenses reveals that the lens with lower optical power (Lens 1) has a greater depth of focus. However, even the short-focal lens (Lens 2) maintains a focus depth of several millimetres. The irregularities observed in the intensity distribution within the rectangular focal spot are not caused by the optical system itself but are a result of the Monte Carlo method used for the simulations. This is evident from the intensity distribution in the initial Gaussian beam simulated using the same method. It should be noted that the parameters for the short-focal lens (right side of **Table 1**) were obtained after additional optimization, deviating slightly from the cylindrical surface geometry to enhance performance.

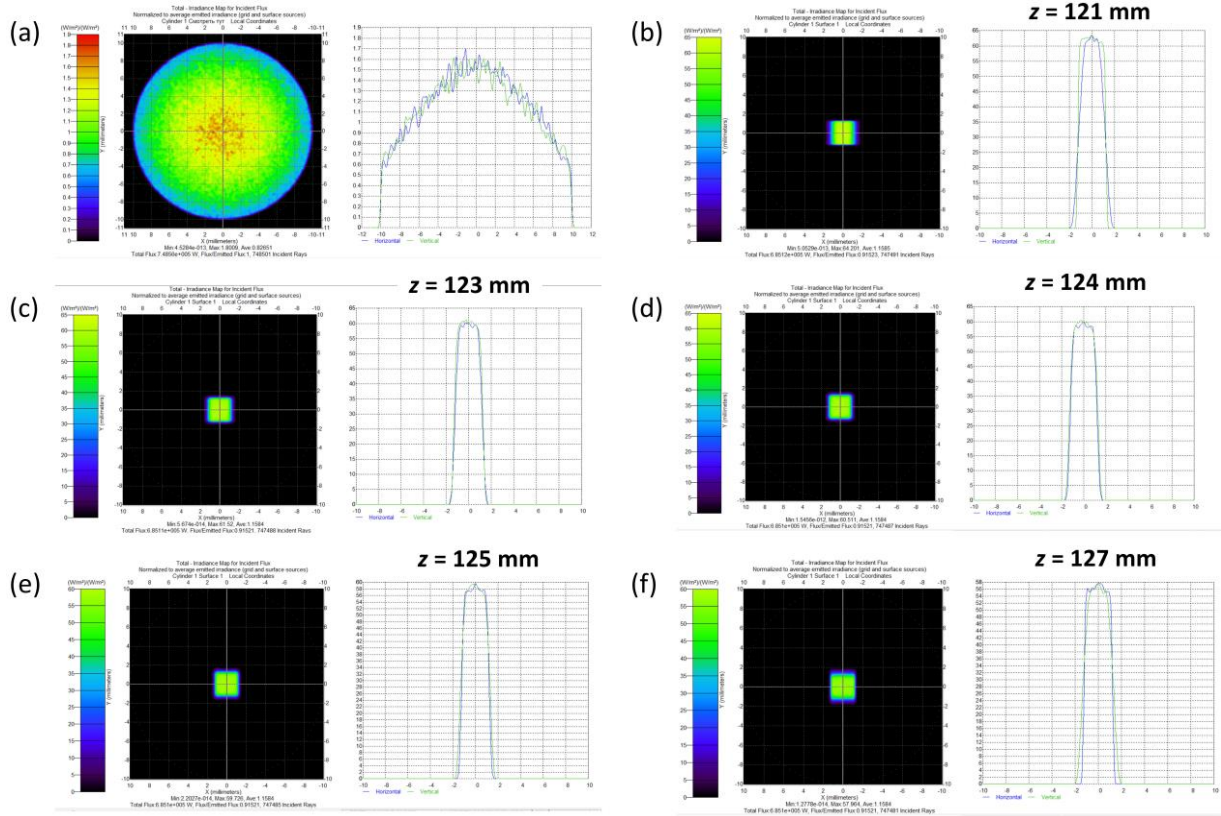

**Figure 10.** Intensity distribution of a laser beam focused by Lens 1 at different distances from the lens: (a) Initial Gaussian intensity distribution obtained in the model with the periphery truncated at the 1/e level by an aperture; (b)–(f) Transverse intensity distribution of the beam at various distances *z* from the lens.

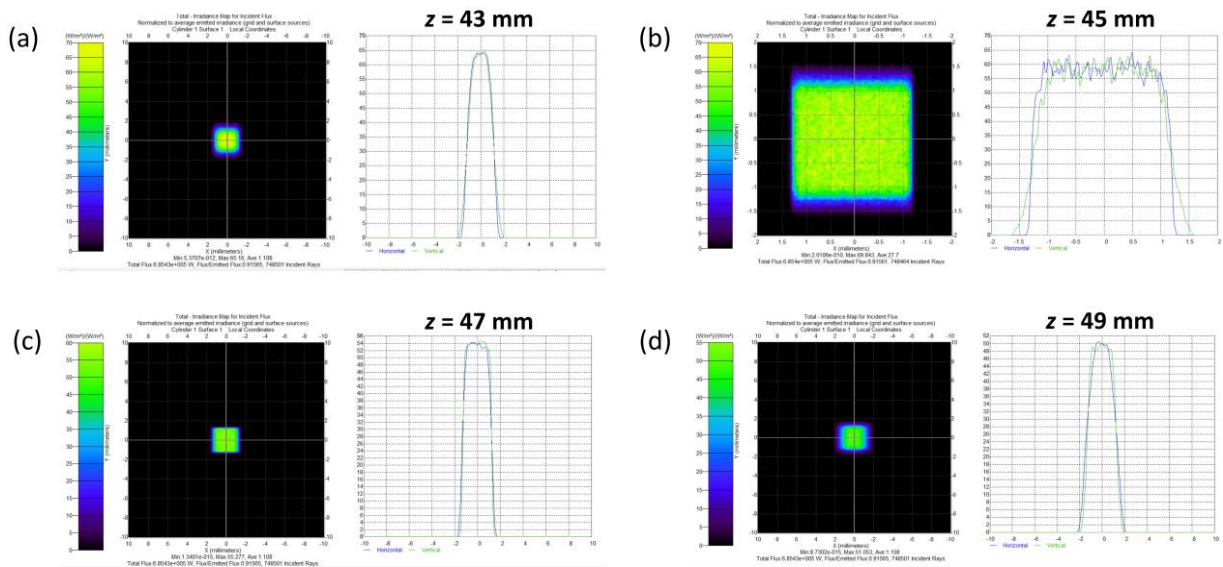

**Figure 11.** Intensity distribution of a Gaussian laser beam (clipped at the level of focus by Lens 2) at various distances from the lens: (b)–(d) show the transverse intensity distribution within the beam at different distances *z*.

**Table 1.** Tilt angles of the flat surfaces of 1D lenses for the long-focal and short-focal combined 2D Fresnel lenses.

| Angle number | Lens 1 ($F = 124$ mm) | | Lens 2 ($F = 45$ mm) | |
|---|---|---|---|---|
| | $\alpha_n$ (°) | $\beta_n$ (°) | $\alpha_n$ (°) | $\beta_n$ (°) |
| 0 | 90 | 90 | 90 | 90 |
| 1 | 87.95 | 87.7 | 84.63 | 83.99 |
| 2 | 85.9 | 85.43 | 79.23 | 78.4 |
| 3 | 83.85 | 83.19 | 73.89 | 73.41 |
| 4 | 81.81 | 81.02 | 68.69 | 69.08 |

## 5. Experimental studies of speckle reduction using the passive method with rectangular prisms of varying heights

### 5.1 Parameters of the experimental setup

Since we were unable to fabricate the required Fresnel lens, we conducted experiments on the effectiveness of the speckle reduction method using an optical system with a similar mechanism for generating sub-beams with varying optical path lengths. This was achieved using rectangular prisms, though the system employs a different optical scheme for achieving uniform illumination. This scheme is based on two rectangular microprism gratings and a Fourier lens, as described in greater detail in [29] (see **Figure 12**) (a similar illumination system was proposed in **[30]**). This optical system is analogous to the proposed design in its speckle reduction mechanism, enabling an evaluation of the speckle reduction efficiency in the suggested configuration.

Rectangular prisms of varying heights were constructed by stacking different numbers of flat, parallel polycarbonate plates with a thickness of $h = 0.6$ mm, resulting in a minimum optical path difference of $\Delta = h \cdot (n - 1)$, where n is the refractive index of polycarbonate. The 2D grating was formed by crossing two 1D gratings at a 90° angle. However, this approach does not ensure an optimal 2D grating structure, leaving certain regions with identical heights. The grating consists of square flat surfaces measuring 0.4 mm, corresponding to the dimensions of microprisms in the 2D microprism grating. The microlenses used in the setup had a focal length of 10 mm. **Table 2** presents the optical path difference matrix of sub-beams generated by our optical illumination system. It indicates that the system can produce no more than 13 decorrelated sub-beams. In the experiment, we used a green laser diode (L520D50), and measurements were conducted at an ambient temperature of 10 °C. The emission spectrum of this laser diode is shown in **Figure 13** [31]. The laser has a spectral width of 2 nm, giving a coherence length of $l = \lambda^2/\Delta\lambda \approx 0.134$ mm, which is 2.63 times smaller than the minimum path difference in the rectangular prism grating. Notably, the laser diode's spectrum at a fixed temperature and operating current is not continuous but comb-like, with peak widths much narrower than the spacing between peaks. The optical setup used in the experiment generates a light beam at each point on the screen with a numerical aperture of $NA_1 = 0.0029$, while the camera observes the screen with a numerical aperture of $NA_2 = 0.000575$ ($NA_1/NA_2 \approx 5$).

**Table 2.** Height profile of the 2D microprism grating (MR) based on two 1D microprism gratings. The black integer values represent plate height differences in units of a single substrate thickness 0.6 mm. The 2D structure has a lateral size of $3.6 \times 3.6$ mm². The red integer values indicate the optical path length differences in the 1D microprism grating structure.

| 12 | 10 | 9 | 7 | 6 | 8 | 9 | 11 | 12 | 6 |
|---|---|---|---|---|---|---|---|---|---|
| 11 | 9 | 8 | 6 | 5 | 7 | 8 | 10 | 9 | 5 |
| 9 | 7 | 6 | 4 | 3 | 5 | 6 | 8 | 18 | 3 |
| 8 | 6 | 5 | 3 | 2 | 4 | 5 | 7 | 8 | 2 |
| 6 | 4 | 3 | 1 | 0 | 2 | 3 | 5 | 6 | 0 |
| 7 | 5 | 4 | 2 | 1 | 3 | 4 | 6 | 7 | 1 |
| 9 | 7 | 6 | 4 | 3 | 5 | 6 | 8 | 9 | 3 |
| 10 | 8 | 7 | 5 | 4 | 6 | 7 | 9 | 10 | 4 |
| 12 | 10 | 9 | 7 | 6 | 8 | 9 | 11 | 12 | 6 |
| 6 | 4 | 3 | 1 | 0 | 2 | 3 | 5 | 6 | |

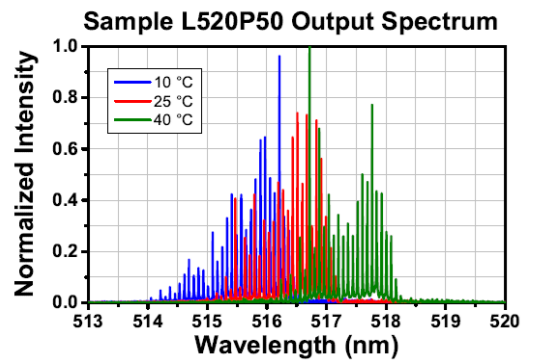

**Figure 13.** Light spectrum of the L520P50 laser diode (LD) at different temperatures [31].

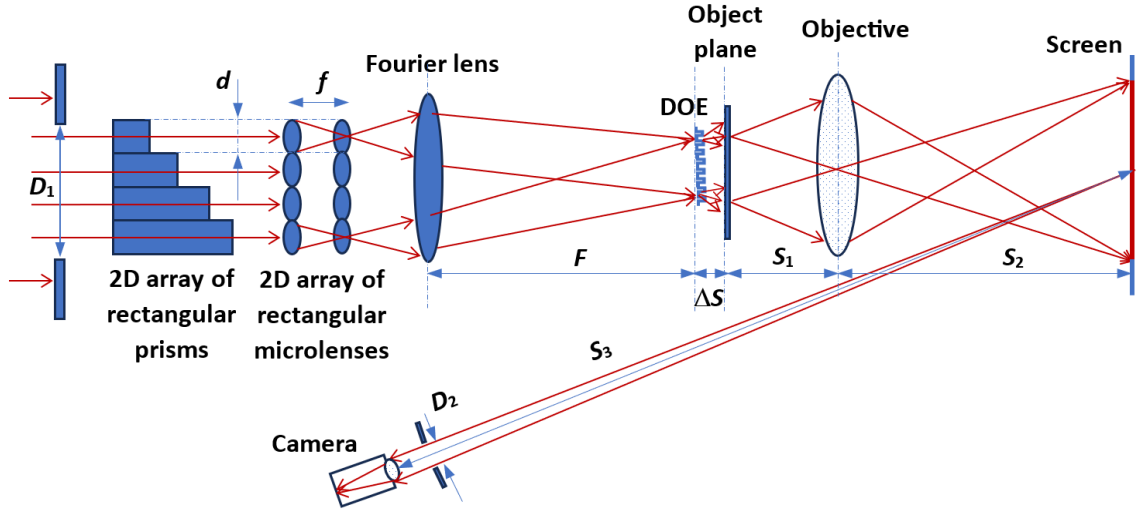

**Figure 12.** The optical setup used to evaluate the effectiveness of the speckle reduction method. Parameters: $D_1 = 4$ mm, $d = 0.4$ mm, $f = 10$ mm, $F = 240$ mm, $\Delta S = 3$ mm, $S_1 = 65$ mm, $S_2 = 185$ mm, $S_3 = 870$ mm, $D_2 = 1$ mm.

*5.2 Results of speckle measurements under different power supply modes*

To evaluate the efficiency of speckle reduction, we conducted a series of speckle measurements using the optical system described above. However, we first measured the speckle pattern produced when the screen was illuminated directly by the laser diode. This allowed us to determine the reduction in speckle contrast solely due to the imperfect temporal coherence of the laser diode. In all measurements, the camera was positioned at a fixed distance from the screen and maintained a constant aperture. The aperture of the objective diaphragm was kept small (1 mm), ensuring that the speckle sizes were much larger than the pixel size of the camera's photodiodes. As a result, there was no significant averaging of the speckle pattern during light intensity integration across the diode area.

When measuring speckles using the proposed optical illumination system, we investigated various laser diode power supply modes. Speckle measurements were conducted under different constant current values as well as under a mode where the current increased linearly from Imin to Imax. The latter mode was chosen to transform the comb-like emission spectrum into a continuous one by averaging over the camera's intensity integration time. This transformation is possible due to the dependence of the laser diode's emission spectrum on both current and temperature. As the current and temperature increase, the spectrum shifts towards the red region, as demonstrated in [32] for GaN-based blue laser diodes. This behavior is not exclusive to GaN diodes, as shown by the temperature-dependent spectrum of our L520P50 laser diode (see **Figure 13**) [31].

The idea behind achieving a continuous spectrum is that, with a sufficient change in current, the spectral shift ensures that each spectral line moves continuously to at least the initial position of a neighboring spectral line in the comb-like spectrum. When the screen was illuminated directly by the laser without using our illumination system, the speckle contrast was measured as $C_1 = 0.42 \pm 0.03$. **Table 3** presents the results of the speckle contrast obtained from the experiments. As shown, the speckle contrast under a constant current remains approximately constant at 0.18 within the experimental error, regardless of the current magnitude. This indicates that the spectral width of the laser diode does not significantly depend on the current within the wide range of currents applied in the experiments. The application of a varying current significantly reduces the speckle contrast to 0.117, which is approximately consistent across three modes with different current variation amplitudes (from 30 to 10 mA). This suggests that the changes in the laser diode's emission spectrum play a significant role, as the resulting outcome in this purely passive method depends solely on the laser diode's spectrum.

*5.3 Analysis of the results and discussion*

Let us examine how specific parameters of the optical setup, the spectrum of the laser diode (LD), and the screen influence the efficiency of speckle reduction. The effect of the emission bandwidth (temporal decorrelation effect) can be represented as the square root of the number $N_1$ of decorrelated scattered light beams resulting from the optical path differences during scattering.

**Table 3.** Results of speckle measurements for different laser diode (LD) power supply modes (current in milliamps).

| N | 1 | | 2 | | 3 | | 4 | | 5 | | 6 | | 7 | |
|---|---|---|---|---|---|---|---|---|---|---|---|---|---|---|
| | $I_{min}$ | $I_{max}$ | $I_{min}$ | $I_{max}$ | $I_{min}$ | $I_{max}$ | $I_{min}$ | $I_{max}$ | $I_{min}$ | $I_{max}$ | $I_{min}$ | $I_{max}$ | $I_{min}$ | $I_{max}$ |
| | 45 | 45 | 55 | 55 | 65 | 65 | 75 | 75 | 45 | 75 | 55 | 75 | 65 | 75 |
| C | 0.18±0.03 | | 0.18±0.03 | | 0.176±0.028 | | 0.18±0.02 | | 0.12±0.02 | | 0.11±0.02 | | 0.12±0.02 | |

Thus, the optical path difference of the scattered light can be determined as the ratio of the square of the speckle contrast for coherent laser illumination to the speckle contrast obtained with the given diode. Taking into account the depolarization of light by the screen, this relationship can be expressed as

$$N_l = \left(0.707 / C_1\right)^2 = \left(0.707 / 0.42\right)^2 = 2.83, \quad (10)$$

thus, the scattered light on the screen exhibits an optical path difference amplitude of 2.83$l$. Previously, we estimated the minimum optical path difference of sub-beams in a rectangular prism array within the decorrelation lengths of our laser beam as 2.63$l$, which is close to the values observed for light scattered on the screen surface. Therefore, for this specific screen and laser, such an optical setup allows for an almost optimal utilization of both temporal and spatial coherence reduction to minimize speckles. However, there are other limitations. Spatial decorrelation cannot achieve a speckle reduction greater than the ratio of the numerical apertures of the objective and the eye, which in our case is about 5. Nevertheless, our optical system provides only $N_2 = 13$ decorrelated laser beams. As a result, the speckle reduction coefficient due to angular diversity cannot exceed $k_\Omega \approx sqrt(13) \approx 3.6$. Consequently, applying this method should lead to a speckle contrast $C_2 \approx C/3.6 = 0.116$, which is significantly lower than the values obtained for the setup with a fixed current. However, it closely matches the values achieved when using the method with a variable current.

This can be explained as follows. Clearly, with a fixed current, we observe a comb-like spectrum with narrow spectral peaks. For the given range of optical path differences [24], the number of decorrelated beams (produced by the screen and our optical system) cannot exceed the number of narrow peaks in the comb spectrum, which in our case is no more than $N_3 = 20$. As a result, the speckle contrast cannot be reduced below $C_2 = 0.707/sqrt(20) \approx 0.158$. This value is slightly lower than the one we obtained using a fixed current, but the difference is minimal. The observed discrepancy can likely be attributed to the presence of spectral peaks with low intensity, which contribute less to the decorrelation process. The agreement between theoretical results and experimental values for the case of a laser diode powered by a variable current (to the third decimal place) demonstrates that a slight amplitude modulation of the current (~15%) during the intensity integration time effectively transforms the comb-like spectrum into a quasi-continuous spectrum. This transformation enables the full utilization of passive speckle reduction methods. Without such modulation, the effectiveness of passive methods is limited by the number of spectral modes in the laser diode's emission spectrum. It is worth noting that such precise agreement between theoretical and experimental results is somewhat unexpected, given that the theoretical model is simplified and does not account for all aspects, such as the non-uniform intensity distribution in the diode's spectrum. Under different conditions, larger discrepancies between theoretical predictions and experimental results should be anticipated. Nonetheless, this level of agreement highlights the robustness of the proposed method under the tested conditions.

## 6. Conclusion

This study highlights critical factors for effective speckle reduction by utilizing the non-ideal temporal coherence and angular diversity of laser light. It demonstrates that achieving a minimum optical path difference between sub-beams exceeding the optical path difference of scattered light is crucial for optimal speckle suppression. The innovative Fresnel lens design, featuring crossed cylindrical Fresnel lenses, efficiently focuses laser illumination into a rectangular spot with high étendue, ensuring uniform and speckle-free illumination for large screens. Additionally, the work addresses the limitations of laser diodes' comb-like spectrum by introducing slight modulation of the drive current, transforming the spectral comb into a continuous spectrum and eliminating constraints on the number of decorrelated sub-beams. These advancements collectively enable efficient speckle reduction regardless of laser spectral modes, providing a robust foundation for developing advanced illumination systems capable of delivering high-quality, speckle-free laser illumination.

**Acknowledgments.** This research was supported by the National Academy of Sciences of Ukraine (grant 0119U001105).